\documentclass[%
 reprint,
superscriptaddress,
showpacs,
 amsmath,amssymb,
 aps,
prl,
]{revtex4-1}

\usepackage{graphicx}
\usepackage{dcolumn}
\usepackage{bm}
\usepackage{graphicx}
\usepackage{dcolumn}
\usepackage{bm}
\usepackage{color}
\usepackage{soul}
\definecolor{blue}{rgb}{0.1,0.7,1.0}
\sethlcolor{blue}
\usepackage{epstopdf}
\usepackage{array}
\usepackage{hyperref}
\usepackage{multirow}

\newcolumntype{d}[1]{D{.}{.}{#1}}

\newcommand{\etal}{\textit{et al.}}
\newcommand{\degc}{$^{\circ}$C}

\newcommand{\tstar}{$T^{\ast}$}
\newcommand{\tmag}{$T_{mag}$}

\newcommand{\pone}{$P\bar{1}$}
\newcommand{\cm}{$C2/m$}

\newcommand{\bcto}{Ba$_{2}$CuTeO$_{6}$}

\newcommand{\bzto}{Ba$_{2}$ZnTeO$_{6}$}

\newcommand{\unit}[2]{#1$\,$#2}
\newcommand{\cax}{\textit{c}-axis}

\newcommand{\bax}{\textit{b}-axis}
\newcommand{\abp}{\textit{ab}-plane}

\newcommand{\jij}{$J_{inter}/J$}
\newcommand{\hp}{$H_{\parallel ab}$}
\newcommand{\hpr}{$H_{\perp ab}$}

\newcommand{\jint}{$J_{inter}$}
\newcommand{\te}{$^{125}\mathrm{Te}$}

\newcommand{\chit}{$\chi(T)$}

\begin{document}

\preprint{APS/123-QED}

\title{$S$=1/2 quantum critical spin ladders produced by orbital ordering in \texorpdfstring{Ba$_{2}$CuTeO$_{6}$}{Ba2CuTeO6} }
\author{A. S. Gibbs}
\email{a.gibbs@fkf.mpg.de}
\affiliation{Max Planck Institute for Solid State Research, Heisenbergstrasse 1, 70569 Stuttgart, Germany}%
\affiliation{Department of Physics, The University of Tokyo, Hongo 7-3-1, Bunkyo-ku, Tokyo, 113-0033 Japan}%
\affiliation{RIKEN Advanced Science Institute, 2-1 Hirosawa, Wako, Saitama 351-0198, Japan}
\author{A. Yamamoto}
\affiliation{RIKEN Advanced Science Institute, 2-1 Hirosawa, Wako, Saitama 351-0198, Japan}
\author{A. N. Yaresko}
\affiliation{Max Planck Institute for Solid State Research, Heisenbergstrasse 1, 70569 Stuttgart, Germany}%
\author{K. S. Knight}
\affiliation{ISIS Facility, Rutherford Appleton Laboratory, Harwell Oxford, Didcot, OX11 0QX, United Kingdom}%
\author{H. Yasuoka}
\affiliation{Max Planck Institute for Chemical Physics of Solids, N\"{o}thnitzer Str. 40, 01187 Dresden, Germany}
\author{M. Majumder}
\affiliation{Max Planck Institute for Chemical Physics of Solids, N\"{o}thnitzer Str. 40, 01187 Dresden, Germany}
\author{M. Baenitz}
\affiliation{Max Planck Institute for Chemical Physics of Solids, N\"{o}thnitzer Str. 40, 01187 Dresden, Germany}
\author{P. J. Saines}
\affiliation{Department of Chemistry, University of Oxford, Inorganic Chemistry Laboratory, South Parks Road,
Oxford OX1 3QR, United Kingdom}%
\author{J. R. Hester}
\affiliation{Bragg Institute, ANSTO, Locked Bag 2001, Kirrawee DC, NSW 2232, Australia.}%
\author{D. Hashizume}
\affiliation{RIKEN Advanced Science Institute, 2-1 Hirosawa, Wako, Saitama 351-0198, Japan}
\author{A. Kondo}
\affiliation{Institute for Solid State Physics, The University of Tokyo, Kashiwa, Chiba 277-8581, Japan}%
\author{K. Kindo}
\affiliation{Institute for Solid State Physics, The University of Tokyo, Kashiwa, Chiba 277-8581, Japan}%
\author{H. Takagi}
\affiliation{Max Planck Institute for Solid State Research, Heisenbergstrasse 1, 70569 Stuttgart, Germany}%
\affiliation{Department of Physics, The University of Tokyo, Hongo 7-3-1, Bunkyo-ku, Tokyo, 113-0033 Japan}%
\affiliation{Institute for Functional Materials and Quantum Technologies, University of Stuttgart,  	Pfaffenwaldring 57, 70569 Stuttgart, Germany}

\date{\today}


\begin{abstract}
The ordered hexagonal perovskite \bcto{} hosts weakly coupled $S$=1/2 spin ladders produced by an orbital ordering of Cu$^{2+}$. The magnetic susceptibility $\chi(T)$ of \bcto{} is well described by that expected for isolated spin ladders with exchange coupling of   \unit{$J\approx{}86$}{K} but shows a deviation from the expected thermally activated behavior at low temperatures below \unit{\tstar{}$\approx25$}{K}. An anomaly in $\chi(T)$, indicative of magnetic ordering, is observed at \tmag{}\unit{$=16$}{K}. No clear signature of long-range ordering, however, is captured in NMR, specific heat or neutron diffraction measurements at and below \tmag{}. The marginal magnetic transition, indicative of strong quantum fluctuations, is evidence that \bcto{} is in very close proximity to a quantum critical point between a magnetically ordered phase and a gapped spin liquid controlled by inter-ladder couplings. 


\end{abstract}

\maketitle



\begin{figure*}[th!]		
\includegraphics[width=2.0\columnwidth]{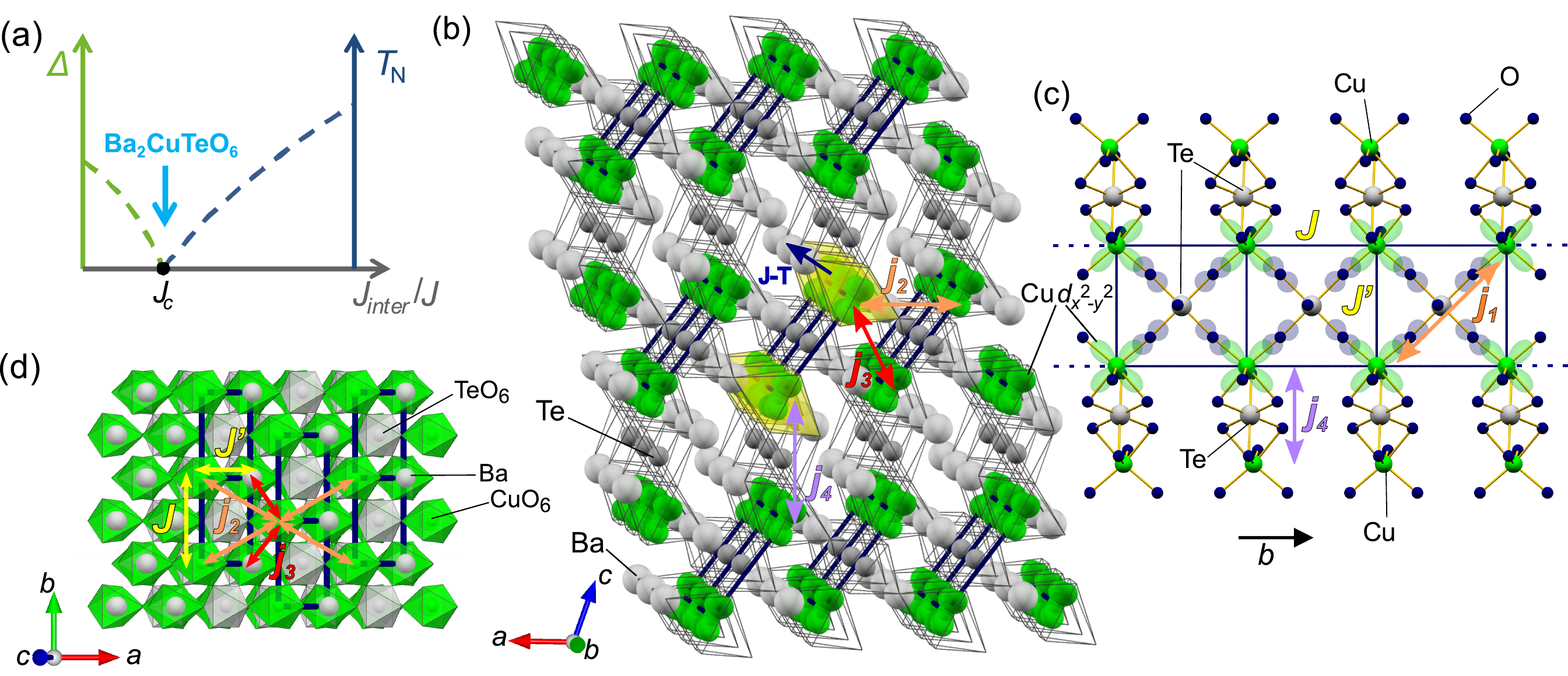}
	\caption{\label{fig:fig1}(Color online) (a) A rough schematic phase diagram at $T=0$ for the two-leg $S=1/2$ antiferromagnetic ladder system with inter-ladder coupling \jint{}. The spin gap $\Delta{}$ is suppressed with increasing inter-ladder coupling strength and goes to zero at the critical coupling $J_c$. For \jint{}$>J_c$ long-range order is expected to develop. The arrow indicates a possible position for \bcto{} in this phase diagram.(b) The chain and ladder arrangement in \bcto{}. The Cu$^{2+}$ chains run along the \bax{} (into the page) and form ladders between layers linked by corner-sharing TeO$_{6}$ octahedra (a single ladder is highlighted in yellow). The ladders form a stacked layer within the \abp{}. The inter-ladder couplings are indicated by arrows. The arrow with label \textquoteleft{}J-T\textquoteright{} indicates the Jahn-Teller distortion direction. (c) The Cu$^{2+}$ chains running along the \bax{} in \bcto{}. These chains can couple to form ladders via superexchange through oxygen ions. The intra-ladder leg and rung exchange couplings are indicated by $J$ and $J^{\prime}$ respectively. The diagonal intra-ladder coupling is labeled $j_1$ and the intra-layer inter-ladder couplings are labeled $j_2$ and $j_3$. The inter-layer coupling between ladders through face-sharing TeO$_{6}$ octahedra is labeled $j_{4}$. (d) The crystal structure of \bcto{} looking onto the \abp{} with the same range as in (b). $J$, $J^{\prime}$, $j_{2}$ and $j_{3}$ are indicated by arrows. The top layer of ladders is shown as thick dark blue lines.}
\end{figure*}

Spin ladder systems have proved to be a rich source of physics over the years. Isolated two-leg $S=1/2$ spin ladders are known to have a spin liquid ground state with a spin gap when both the rung coupling $J_{r}$ and leg coupling $J_{l}$ are non-zero \cite{Barnes1993}. When couplings between the ladders, \jint{},  are present the system is pushed towards long-range ordering. With increasing \jint{}, a quantum critical point (QCP) appears at the onset of long-range ordering \jint{}=$J_{c}$, as depicted in the schematic phase diagram shown in Fig. \ref{fig:fig1}(a). This is the case for both 3D coupled ladders \cite{Normand1996,Troyer1996}, a 2D planar array of ladders \cite{Normand1996,Imada1997}  and a 2D ‘stacked’ array of ladders \cite{Johnston2000,Dalosto2000}. The critical inter-ladder coupling is of the order \jij{}$\approx$0.1-0.4 when $J=J_{l}=J_{r}$ but the exact value depends not only on the the geometry of inter-ladder coupling but also the ratio of $J_{r}/J_{l}$ and the additional interactions in the system. 

Despite the presence of quantum criticality in ladders being well established theoretically, the experimental realization still remains challenging. Very few materials have so far been found to be near to this QCP. Often the candidate compounds turn out to be in the strong-leg limit, for example bis(2,3-dimethylpyridinium)tetrabromocuprate \cite{Shapiro2007}, or the strong-rung (dimer) limit such as (C$_{5}$H$_{12}$N)$_{2}$CuBr$_{4}$  \cite{Thielemann2009} and CaV$_{2}$O$_{5}$ \cite{Korotin2000,Ohama2001}, which places those systems far from ladder physics. It is not easy to tune \jint{} close to $J_{c}$.  (Dimethylammonium)(3,5-dimethylpyridinium)CuBr$_{4}$ was shown to contain isotropic ladders with a strong antiferromagnetic inter-ladder coupling \jij{}=0.32, placing it close to the QCP, and displays long-range magnetic ordering at \unit{2}{K} \cite{Awwadi2008,Hong2014}.  The observation of a pronounced specific heat anomaly at the magnetic transition indicates that the inter-ladder coupling is too strong to capture a clear signature of the criticality. 

Orbital ordering often reduces the magnetic dimensionality of systems \cite{Khomskii2005} due to the anisotropic overlap of orbital wave functions. In La$_{2}$RuO$_{5}$ the orbital order leads to the formation of spin ladders on a quasi-2D crystal structure \cite{Eyert2006}. Similarly, in CuSb$_{2}$O$_{6}$, quasi-1D chains are formed via orbital ordering \cite{Kasinathan2008}. Since the low dimensionality is provided by orbital ordering while the crystal structure remains quasi-2D or 3D, remnants of 2D and 3D interactions are often not negligibly small and could supply the necessary inter-ladder coupling to approach the QCP. 

 \bcto{} with $S$=1/2 Cu$^{2+}$ crystallizes in an ordered hexagonal perovskite structure \cite{Kohl1974}. As can be seen in Figures \ref{fig:fig1}(b) and (d), the crystal structure consists of alternate stacking along the \cax{} of layers with triangular arrangements of CuO$_{6}$ octahedra and layers with triangular arrangements of TeO$_{6}$ octahedra. The CuO$_{6}$ octahedra and the TeO$_{6}$ octahedra in neighboring layers are connected alternately by their corners and faces, giving rise to the unit cell containing two Cu-layers and two Te-layers along the \cax{}. Since Cu$^{2+}$ is Jahn-Teller active, uniform Jahn-Teller distortion of the CuO$_{6}$ octahedra (Fig. \ref{fig:fig1} (b)) occurs below \unit{$T_{J-T}\approx850$}{K}, which leads to the emergence of ferro-orbital ordering of the $d_{x^2-y^2}$ orbitals \cite{Khomskii2014}. This should make the magnetic coupling of the Cu$^{2+}$ $S$=1/2 spins anisotropic. Hybridization of the $d_{x^2-y^2}$ orbitals with 2$p$ orbitals of the four oxygen ions lying in the plane of the $d_{x^2-y^2}$ orbital, and hopping between O $p$ states within the oxygen square planes of the corner-linked TeO$_{6}$ octahedra are anticipated. The superexchange process using these hybridisations highly likely results in stronger coupling of $S$=1/2 Cu$^{2+}$ within the corner sharing CuO$_{6}$-Te-CuO$_{6}$ layered unit (Fig. \ref{fig:fig1}(b) and (c)). As can be seen in Fig. \ref{fig:fig1}(c), the two couplings, $J$ between the neighboring Cu$^{2+}$ along the \bax{} within the triangular plane and $J^{\prime}$ between the pairs of Cu$^{2+}$ in adjacent Cu planes, should be dominant, which would give rise to the formation of $S$=1/2 ladders.

In this study we show that the magnetic susceptibility and NMR data on \bcto{} are fully consistent with the presence of coupled $S$=1/2 ladders. The sizeable inter-ladder coupling places this compound almost exactly at the QCP. A marginal magnetic transition, lacking clear signatures of a transition in NMR, specific heat and neutron diffraction data, was identified in the magnetic susceptibility $\chi(T)$ which we argue to mirror the presence of strong quantum fluctuations around the QCP.


Polycrystalline samples were prepared from stoichiometric mixtures of dried BaCO$_{3}$, CuO and TeO$_{2}$ under flowing oxygen at 950-1000\degc{}. Powder x-ray diffraction and neutron diffraction on HRPD at ISIS \cite{Ibberson1992,Ibberson2009} confirmed the samples to be of single phase. Single crystals were grown with a flux method based upon that of K\"{o}hl \etal{} \cite{Kohl1972,Kohl1974} using BaCO$_{3}$, CuO and TeO$_{2}$. Single crystal x-ray diffraction at \unit{$T=296$}{K} confirmed the structure to be consistent with that reported by K\"{o}hl \etal{} \cite{Kohl1974} but in a higher symmetry space group of $C2/m$ \footnote{We note that high resolution neutron diffraction on HRPD at ISIS showed the presence of a very weak transition to \pone{} symmetry at  \unit{$T=287$}{K}. The details will be reported elsewhere (A. S. Gibbs, K. S. Knight, P. J. Saines and H. Takagi, in preparation)}{}. In \chit{} for the single crystals and powders no Curie-like contribution could be well fitted. A best estimate of the magnetic impurity content indicates $<$0.1\% $S$=1/2 impurities, showing that our samples are extremely clean, partly due to full cation ordering caused by the contrasting ionic radii and valences of Cu$^{2+}$ and Te$^{6+}$. Magnetization and specific heat were measured using a Quantum Design MPMS and PPMS respectively. \te{} NMR ($I$=1/2 and about 7\% natural abundance) measurements at a fixed frequency of \unit{55.44}{MHz} (corresponding to  $\mu_{0}H=4.12\,$T)  were performed for \unit{2}{K}$\leq{}T\leq{}300\,$K using a conventional pulsed NMR technique. LSDA+U calculations were performed for the \cm{} structure \cite{LAZ95,YAF03}. 


\begin{figure}[hbt]		
\includegraphics[width=1\columnwidth]{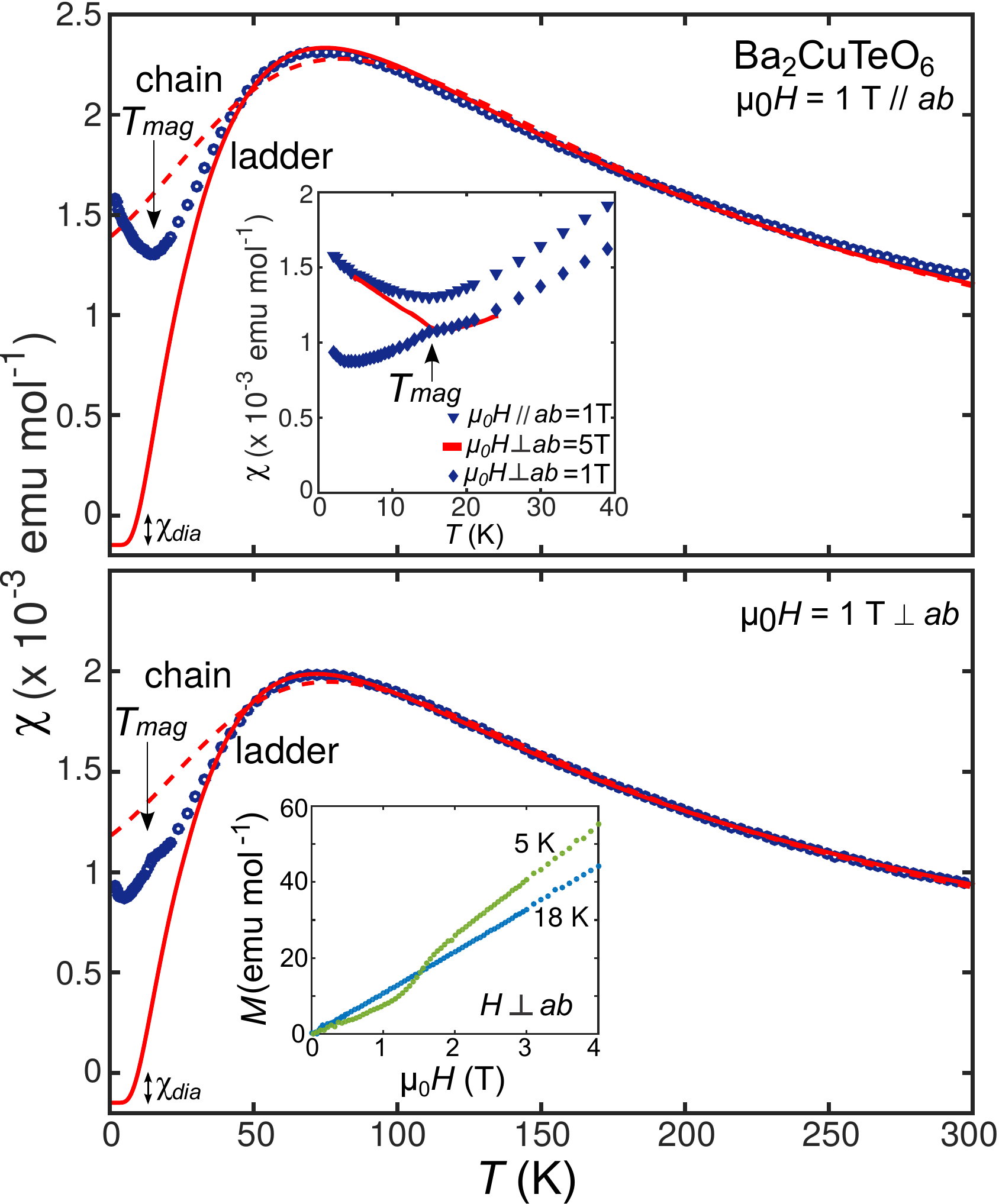}
	\caption{\label{fig:fig2} (Color online)  \chit{} for \bcto{} single crystal arrays with the magnetic field of \unit{$\mu_0H=1$}{T} applied parallel (upper panel) and perpendicular (lower panel) to the \abp{}. The dashed and solid lines indicate chain and ladder model fits respectively. The inset to the upper panel shows the low temperature region for both orientations along with the high field behavior for \hpr{}. The inset to the lower panel shows $M(H)$ of \bcto{} for \hpr{}.}
\end{figure}

\begin{figure}[hbt]		
\includegraphics[width=1.0\columnwidth]{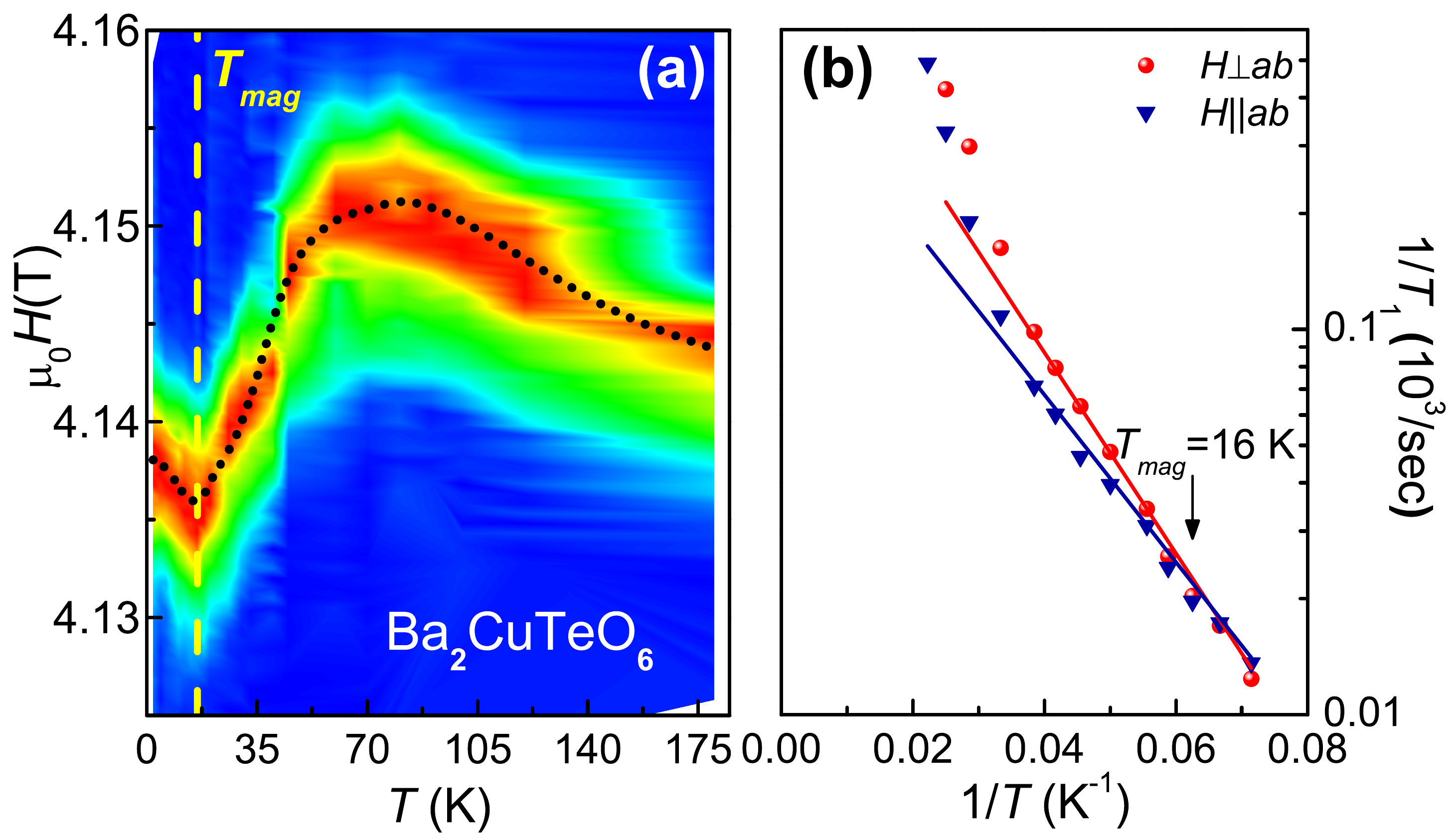}
	\caption{\label{fig:fig3}(Color online) (a) Contour plot of the spin echo intensity of the \te{}-field-sweep NMR spectra as a function of temperature (NMR frequency: 55.44 MHz) for a polycrystalline sample. The dotted line is a guide to the eye. (b) Semi-log plot of $1/T_{1}$ versus 1/$T$. Solid lines correspond to the behavior $1/T_{1}$ =$ A$exp(-$E_{a}/k_{\mathrm{B}}T$) at high temperatures.}
\end{figure}

Examination of the temperature dependence of $\chi(T)$, shown in Fig. \ref{fig:fig2} confirms the hypothesis of \bcto{} as a quasi-1D ladder system. The broad overturn around \unit{$T\approx75$}{K} is characteristic of low-dimensional systems.  Fits to the inverse susceptibility for \unit{170}{K}$\leq{}T\leq{}300\,$K gave $\theta_{W}\approx{}-113\,$K indicating reasonably strong antiferromagnetic interactions and an effective moment of ~1.96$\mu_{B}$/Cu$^{2+}$. $\chi(T)$ is not particularly well fitted by a Heisenberg chain model \cite{Bonner1964,Hatfield1980} but it is better described by an isolated two-leg ladder model \cite{Johnston2000} above \unit{$T=35$}{K} as can be seen in Fig. \ref{fig:fig2} (expressions are given in the Supplemental Material (SM) \cite{SM}). The ladder model fit, including correction for core diamagnetism $\chi_{dia}=-1.48\mathrm{x10^{-4} emu\,mol^{-1}}$, for \hpr{} gives \unit{$J/k_{\mathrm{B}}\approx\,86$}{K}, $J^{\prime}/J\approx\,0.98$, and $g\approx\,2.08$, with \unit{$\Delta/k_{\mathrm{B}}\approx\,40$}{K}. At low temperatures $\chi(T)$ remains finite and deviates from ladder behavior as will be discussed later.

The \te{} NMR data supports the spin ladder picture discussed for the $\chi(T)$ data. The \te{} NMR  spectrum of the powder sample is the sum of two components with a 1:1 ratio in agreement with the two inequivalent Te sites present in the crystal structure (see Fig. \ref{fig:fig1}). Each spectrum could be consistently fitted using an anisotropic Knight shift tensor. The color contour plot of the \te{} NMR intensity in the temperature-magnetic field plane, shown in Figure \ref{fig:fig3}(a), represents the temperature dependence of the Knight shift, which agrees well with  \chit{}. The temperature dependence of the spin-lattice relaxation rate ($1/T_{1}$) for both field directions (\hpr{} and \hp{}) for arrays of single crystals can be fitted reasonably well with an Arrhenius type of behavior ($1/T_{1}$ = $A$exp($-E_a/k_{\mathrm{B}}T$)) at high temperature above \unit{14}{K}, as shown in Figure \ref{fig:fig3}(b), giving an activation energy \unit{$E_{a}/k_{\mathrm{B}}\approx50$}{K},  close to the expected spin gap $\Delta$ in the \textquoteleft{}pure\textquoteright{} spin ladder limit extracted from \chit{}. 

Estimates of effective exchange coupling constants from LSDA+U band structure calculations support the previously described ladder picture. The total energy $E(\mathbf{q}, \phi{})$ was calculated as a function of a wave vector $\mathbf{q}$ and of an angle $\phi{}$ between spins of two Cu$^{2+}$ ions in the \unit{$T=300$}{K} monoclinic \cm{} unit cell for a number of spin-spiral structures. Effective exchange coupling constants were evaluated by fitting $E(\mathbf{q}, \phi{})$ to a classical Heisenberg model. We obtain antiferromagnetic leg coupling \unit{$J$=33.8}{meV} and rung coupling \unit{$J^{\prime}$=33.0}{meV} for \unit{$U=5$}{eV}, yielding almost isotropic $J^{\prime}/J\approx0.98$ in excellent agreement with the $\chi(T)$ fit.  There is roughly a factor of four difference between $J$ obtained experimentally and that from the \unit{$U=5$}{eV} calculation. The calculated $J$ parameters are found to scale down with increasing $U$ approximately as $1/U$ and the agreement with the experimental values are not unreasonable given the uncertainties present. As expected, the other parameters are smaller than $J$ and $J^{\prime}$. The diagonal coupling $j_{1}$ is antiferromagnetic with $j_{1}/J\approx0.05$. There are three main inter-ladder couplings (see Fig. \ref{fig:fig1} (b-d)). $j_{2}$ and $j_{3}$ are between ladders in the same plane (parallel to the \abp{}) whereas $j_{4}$  couples ladders between planes through the face-shared TeO$_{6}$ octahedra (Fig. \ref{fig:fig1}(b-c)). $j_{2}$ and $j_{3}$ are ferromagnetic with $j_{2}/J\approx{}-\,0.06$ and $j_{3}/J\approx{}-\,0.05$. The effects of $j_{2}$ and $j_{3}$ are likely very much reduced due to the frustrated two bond geometry. The \textquoteleft{}inter-plane\textquoteright{} coupling $j_{4}$ appears to be the most important and is antiferromagnetic with $j_{4}/J\approx0.03$.  These inter-ladder couplings are small but not negligible, with a magnitude of the order of 0.1$J$, which may potentially bring the system towards the QCP. 

\begin{figure}[t]		
\includegraphics[width=1.0\columnwidth]{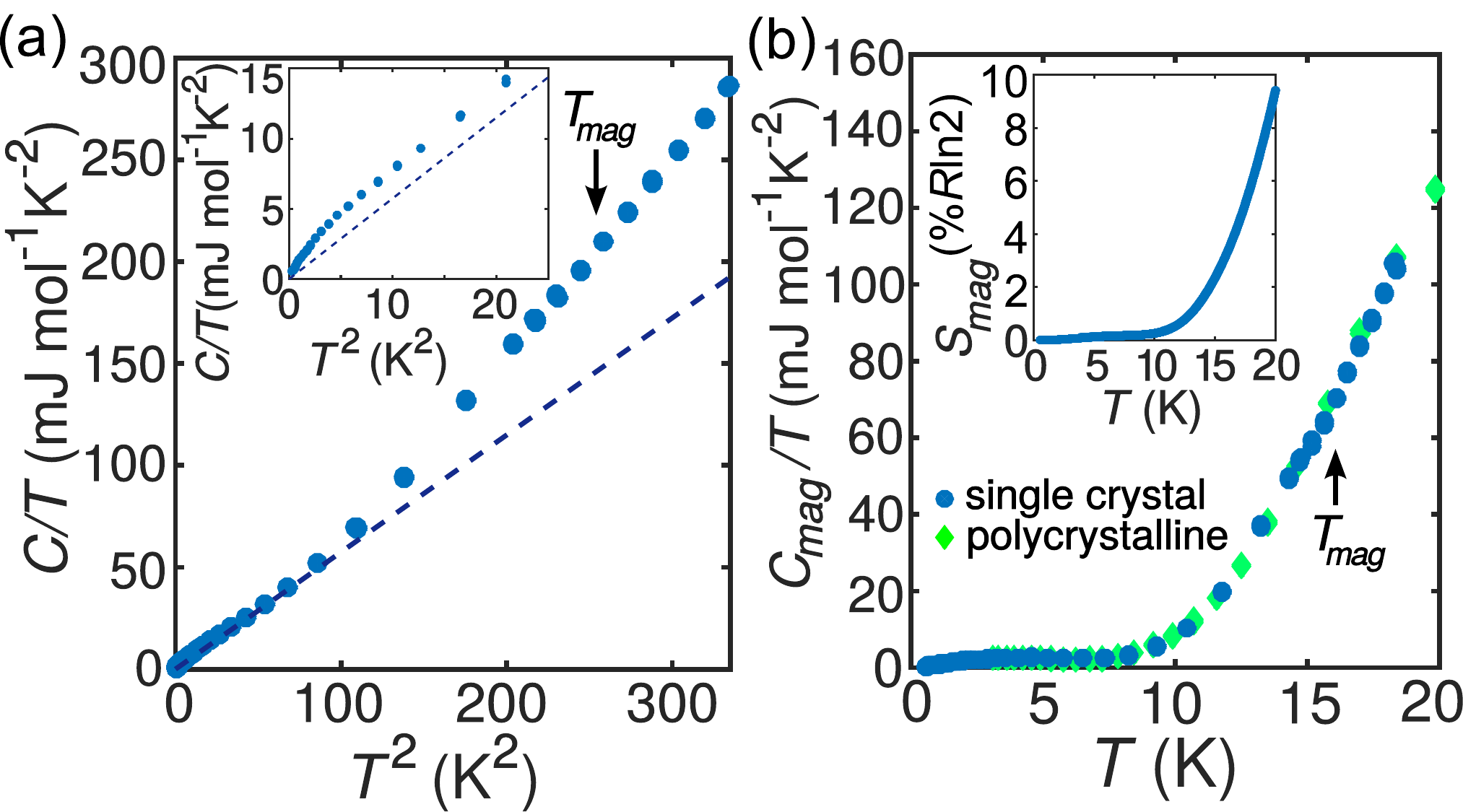}
	\caption{\label{fig:fig4} (Color online) (a) The specific heat $C$ of single crystal \bcto{} plotted as $C/T$ vs. $T^2$. The dashed line indicates an approximate lattice $T^3$ term as described in the text. The inset shows an enlarged view of the low temperature region. (b) The estimated magnetic specific heat $C_{mag}(T)$ for \bcto{} plotted as $C_{mag}/T$ vs. $T$.  The inset shows the estimated magnetic entropy $S_{mag}(T)$ as a percentage of $R$ln2 up to \unit{$T=20$}{K}.}
\end{figure}

\chit{} deviates strongly from the expected ladder behavior below approximately \unit{\tstar{}$\approx$25}{K}, and at \unit{\tmag{}$\approx$16}{K} a cusp is seen below which the anisotropy of \chit{} becomes profound (inset to upper panel of Fig. \ref{fig:fig2}). With lowering temperature below \unit{\tmag{}$\approx$16}{K},  $\chi_{\parallel ab}(T)$ shows a weak increase but in contrast $\chi_{\perp ab}(T)$ shows a decrease. Note that the upturn below \unit{\tmag{}$\approx$ 16}{K} for $\chi_{\parallel ab}(T)$ is reproduced by the temperature dependence of the \te{} NMR intensity (Fig. \ref{fig:fig3}(a)), confirming the intrinsic origin of this behavior \footnote{The NMR data were taken at a frequency of \unit{55.44}{MHz}, corresponding to a magnetic field of \unit{4.12}{T} and in this high field region the susceptibility for \hpr{} also increases below \tmag{} therefore the polycrystalline samples show an increase in \chit{} below \tmag{} at this field since both $\chi$(\hp) and $\chi$(\hpr) increase below \tmag{}.}. The cusp is seen reproducibly in all batches of single crystals and all polycrystalline samples (see SM \cite{SM}). In the magnetization $M-H$ curve (inset to lower panel of Fig. \ref{fig:fig2}), a rapid increase, almost step-like, was observed around \unit{1.5}{T} only for \hpr{}.  Since the behavior of \chit{} well above \unit{1.5}{T} is quite similar to $\chi_{\parallel ab}(T)$, this anomaly likely represents a spin reorientation transition. All of these results point to the conclusion that the anomaly at \unit{\tmag{}$\approx$ 16}{K} is due to magnetic ordering. We argue that the ordering is associated with the small but finite inter-ladder coupling and that the system has marginally passed the expected QCP. The deviation of \chit{} from ladder behavior at low temperatures could be understood as the manifestation of the effect of inter-ladder coupling and quantum criticality. 

Despite this weak but clear feature in \chit , specific heat measurements in both zero field and applied fields of \unit{5}{T}  and \unit{13}{T} failed to find any clear anomaly indicative of ordering at \tmag{} (see Fig. \ref{fig:fig4}(a) and SM \cite{SM}).  Neutron diffraction experiments using a large-volume powder sample on Echidna at the OPAL reactor \cite{Liss2006} showed no magnetic Bragg peaks. No anomaly in the NMR spin-lattice relaxation rate, $1/T_{1}$, measured on aligned single crystals was detected upon passing through \tmag{} (Fig. \ref{fig:fig3}(b)). This absence of clear signals of long-range ordering indicates that the cusp seen in susceptibility is a \textquoteleft{}marginal\textquoteright{} phase transition.  We argue that the marginality arises from the quantum fluctuations around the QCP.

The described lack of specific heat anomaly at the marginal phase transition implies that the peak is smeared out by quantum critical fluctuations and indeed we detect indications of unquenched magnetic entropy at low temperatures. As seen in the inset to Fig. \ref{fig:fig4}(a), $C/T$ vs. $T^2$ is not linear, even below \unit{5}{K}. A best estimate of the phononic $T^3$ lattice contribution is indicated as a dashed line in Fig. \ref{fig:fig4}(a). For the nonmagnetic analogue \bzto{},  $C/T$ vs. $T^2$ is completely linear below \unit{5}{K} \footnote{\bzto{} has a higher symmetry structure than \bcto{} in the low temperature regime (with space group \cm{} compared to the \pone{} space group of \bcto{} in this range [21]) and can therefore not be used for a reliable subtraction. However, the similar Debye temperature indicates that our estimated $T^3$ term is reasonable.} and the Debye temperature of  \bcto{} ($\theta_D\approx\,323\,\mathrm{K}$), estimated from the dashed line in Fig. 4, agrees reasonably well with that of \bzto{} ($\theta_D\approx\,377\,\mathrm{K}$).  These facts indicate that the dashed line in Fig. \ref{fig:fig4}(a) is a reasonable estimate of the lattice contribution and that there is an additional contribution to the specific heat other than the lattice which can naturally be ascribed to $S$=1/2 spins. The magnetic specific heat, $C_{mag}(T)$ was estimated by subtracting the aforementioned phononic $T^3$ term from the total specific heat, the result is shown in Fig. \ref{fig:fig4}(b). $C_{mag}(T)$ was then integrated to provide an estimate of the magnetic entropy  $S_{mag}(T)$, shown in the inset to Fig. \ref{fig:fig4}(b), which is approximately 4.5\% of $R$ln2 at \unit{17}{K}. Despite the presence of such a large magnetic entropy at low temperature, no clear anomaly is present in $C_{mag}(T)$ at \unit{\tmag{}$\approx16$}{K} and the entropy is quenched only gradually to \unit{$T=0$}{K}. We argue that the large magnetic specific heat observed around \unit{\tmag{}$\approx$ 16}{K} represents the quantum fluctuations associated with the close proximity to the QCP.

In conclusion, we have discovered that \bcto{} is a clean, site ordered system with `hidden' antiferromagnetic $S$ =1/2 spin ladders due to ferro-orbital ordering on the Cu$^{2+}$ site. The very weak signature of order in \chit{}, without any corresponding anomaly in specific heat, neutron diffraction or NMR points to the conclusion that \bcto{} is almost precisely at the QCP for the antiferromagnetic S =1/2 weakly coupled ladder system, where quantum fluctuations dominate. Compared to other systems close to the QCP, such as (Dimethylammonium)(3,5-dimethylpyridinium)CuBr$_4$, in which clear thermodynamic signatures of order are seen, \bcto{} may provide a better opportunity to study the quantum criticality of the weakly coupled ladder system.

\begin{acknowledgments}
The authors thank J\'{e}r\^{o}me Dufour, George Jackeli, Fr\'{e}d\'{e}ric Mila,  Masaaki Nakamura, Seiji Niitaka and Andreas W. Rost for invaluable discussions. A.S.G. thanks Prof. A. P. Mackenzie and Prof. P. Lightfoot for support in the early stages of the project. This work was partly supported by Grant-in-Aid for Scientific Research (S) (Grant No. 24224010) and the Alexander von Humboldt foundation.
\end{acknowledgments}

\end{document}